\documentstyle[pra,aps,preprint]{revtex}
\tightenlines

\begin{document}

\title{Local-field approach to the interaction of an ultracold dense Bose gas
 with a light field}
 
\author{Konstantin V. Krutitsky
		\thanks{
		   Permanent address: 
		   Ulyanovsk Branch of Moscow Institute of Radio 
		   Engineering and Electronics of Russian Academy of Sciences, 
		   48, Goncharov Str., Ulyanovsk 432700, Russia. 
		 },
	Frank Burgbacher
        \thanks{e-mail: frank.burgbacher@uni-konstanz.de},
 	and J\"urgen Audretsch 
		\thanks{e-mail: Juergen.Audretsch@uni-konstanz.de}
} 
\address{Fakult\"at f\"ur Physik, 
		 Universit\"at Konstanz, Fach M 674, 
		 D-78457 Konstanz, Germany 
}

\date{\today}

\maketitle

\begin{abstract}
The propagation of the electromagnetic field of a laser through a dense Bose
gas is examined and nonlinear operator equations for the motion of the center
of mass of the atoms are derived. The goal is to present a self-consistent set of coupled Maxwell-Bloch equations for atomic and electromagnetic fields generalized to include the atomic center-of-mass motion. Two effects are considered: The ultracold gas forms a medium for the Maxwell field which modifies its propagation properties.
Combined herewith is the influence of the dipole-dipole interaction between
atoms which leads to a density dependent shift of the atomic transition 
frequency. It is expressed in a position dependent detuning and is the reason 
for the nonlinearity. This results in a direct and physically transparent way 
from the quantum field theoretical version of the local-field approach to 
electrodynamics in quantum media. The equations for the matter fields are 
general. Previously published nonlinear equations are obtained as limiting 
cases. As an atom optical application the scattering of a dense beam of a Bose 
gas is studied in the Raman-Nath regime. The main conclusion is that for 
increasing density of the gas the dipole-dipole interaction suppresses or enhances the scattering depending on the sign of the detuning.
\end{abstract}


\section{Introduction}

During the last decade we have witnessed a rapid progress in the field of atom 
optics \cite{MLY92}. Examples for experimental success are Bragg scattering of 
an atomic beam from a standing wave\cite{MAR88}, reflection of atoms from an 
evanescent wave\cite{FER94}, and interference of atomic beams\cite{RIE91}. 
Theoretically predicted are among other things the transmission and reflection 
of atoms from standing light fields\cite{ZHA94b}, tunneling\cite{TRI96}, and 
also it seems to be possible to focus atomic waves with the help of 
inhomogeneous fields\cite{ROH96}. All these results refer to a physical regime in which the system can be described effectively as a single particle system. The respective quantum equations of motion contain no interaction between particles, which makes them linear.

With the experimental realization of Bose condensates\cite{AND95,BRA95,DAV95,AND96}, atom optics has 
reached a new stage of development. The influence of many body effects 
on the atomic motion in light fields became important, demanding a quantum 
field theoretical treatment of matter. In addition, in the presence of 
light the polarizable atoms are subject to the resonant dipole-dipole 
interaction  which causes the equation of motion to become nonlinear. Perhaps 
the most well known effect of nonlinear wave propagation is the formation of 
solitons, which have also been studied for condensates in the pioneering 
theoretical work published in \cite{LEN93,LEN94,ZHA94}. For the shaping of 
these solitons see \cite{HOL97}. In the framework of this new domain of 
nonlinear atom optics too, atom optical devices were studied \cite{CHO97}. 
For beam splitters for condensates based on the Bragg scattering of a 
coherent beam see \cite{ZHA94a,ZHA95,SCH95}. The underlying theoretical 
scheme has been criticized in \cite{CAS95} and \cite{WAL97}, mainly with 
regard to the treatment of the electric field inside the atomic sample of 
polarizable atoms. Treating the dipole-dipole interaction as a contact 
interaction is regarded as a too severe approximation. In addition, an 
approximation restricting to low densities of the condensate is made.

To overcome these problems, more rigorous approaches to describe the joint 
dynamics of atoms and light in the quantum regime have been developed. First 
without the inclusion of atomic motion\cite{BOW93}. Later, with applications 
to atom optics in mind\cite{CAS95}, generalized Maxwell-Bloch equations were 
derived which in principle include retardation and therefore correctly 
describe dispersion, as well as the local-field effect, and accordingly give 
the correct Maxwell equations. In \cite{MOR95} the formalism was used to 
calculate the refractive index of a dilute Bose gas and in \cite{WAL97} 
atoms moving in an evanescent wave and the gravitational field were considered. We want to contribute to this field of research.

The second subject to which this paper is related are the local-field effects. According to the Lorentz-Lorenz relation the local electric field in a medium differs from the macroscopic field. This leads to the local-field effect in classical, quantum and nonlinear optics which manifests itself mainly if the density of the molecules, which constitute the medium, is high. In classical optics the Lorentz-Lorenz relation leads to the more general Clausius-Mossotti formula for the refractive index of the medium in comparison with that based on the standart response theory. In quantum optics the local-field effect leads to an essential modification of the lifetime of the atom embedded in a dense dielectric medium\cite{GLA91,BAR92}, to statistical effects in superfluorescence and amplified spontaneous emission\cite{RAI92}. Many new interesting phenomena caused by the local-field effect can be found in nonlinear optics, where the local-field correction gives rise to the nonlinearity in Bloch equations and as a consequence to the optical bistability that is intrinsic to the material and doesn't require an external feedback (see, for example,\cite{BEN86,FRI89,ING90,CRE96} and references therein), to intrinsic self-phase modulation in self-induced transparency\cite{STR88,BOW91}, to linear and nonlinear spectral shifts\cite{FRI73,FRID89}, to great enhancement of the index of refraction with no absorption in the systems exhibiting coherence-induced lasing without inversion\cite{DOW93}, to ultrafast intrinsic optical switching in a dense medium of two-level atoms\cite{CRE92} and to a nonlinear transparency of a thin resonant layer remaining near to its ground state\cite{BENEDICT91}.

In the present paper we shall show that the local-field effect plays also an important role in the atom optics of ultracold Bose gas. It causes the nonlinearity in the Bloch equations for the atomic quantum field. This leads in its turn to the nonlinear phenomena in atom optics.
Our purpose is twofold: With regard to the development of a general scheme we study in a first part an electromagnetic field propagating in an ultracold gas (Bose condensate) of neutral polarizable atoms. Our intention is to give 
a straightforward and short derivation of the coupled dynamical equations 
for the Maxwell field and the matter field operators taking into account the atomic center-of-mass motion. The resulting equations form the basis 
of the respective nonlinear atom optics in the domain, where the resonant 
dipole-dipole interaction is the dominant effect. Starting from the 
Hamiltonian in second quantization we present a physically intuitive scheme 
which reflects the fact that we are dealing with electrodynamics in media. 
We are thereby directly led to the related concept of the local-field approach. The Lorentz-Lorenz formula and the Clausius-Mossotti relation follow immediately, so that the optical properties can be discussed. The origin of the 
nonlinearity of the matter field equation becomes transparent. Important is 
also, that no approximation with regard to the density will be made. 
Previously considered equations can be obtained as limiting cases for low 
density. It is not necessary to refer to a particular quantum state of the gas.

The second purpose of the paper is to apply the general results obtained for 
the dynamics of the field operators to a particular device. In principle we 
have to solve a coupled system of equations. Under certain assumptions it is 
possible to solve the Maxwell equations without restricting the density of 
the gas. We study the scattering of the atoms from a standing light wave in 
the strong coupling regime. The dependence on the density of the gas is 
discussed.

\section{The model}

We discuss the combined system of a Bose gas of identical neutral 
polarizable atoms interacting with an incident monochromatic field 
${\bf E}_{in}({\bf r},t)$ with frequency $\omega_L$. The probe field 
${\bf E}_{in}$ is a classical external field. No operator is related with it. 
It is fixed and therefore not subject to a dynamics. For example, it may be 
a standing or running laser field. The quantized center-of-mass motion of 
the atoms is taken into account. The atoms are represented as point-like 
neutral two-level atoms with states $|1\rangle$ and $|2\rangle$ with ``bare" 
transition frequency $\omega_a$. The atoms interact via the resonant 
dipole-dipole interaction of the induced point dipoles. This interaction is 
mediated by the exchange of {\em photons} with annihilation and creation 
operators $\hat{c}_{{\bf k}\lambda}$ and $\hat{c}^{\dagger}_{{\bf k}\lambda}$ 
referring in lowest order to plane waves in vacuum with wave vector ${\bf k}$ 
and polarization $\lambda$.

The total atom-field Hamiltonian may then be written in the electric dipole 
approximation in the length gauge or ${\bf d}{\bf E}$ coupling~\cite{COH89} 
in the form

\begin{eqnarray}\label{ham-many-part}
H &=& H_A + H_F + H_{AI} + H_{AF}
\;,
\nonumber\\
H_A &=& \sum_i \left\{- \frac{\hbar^2 \nabla_i^2}{2m} + \hbar\omega_a 
\sigma_i^\dagger\sigma_i \right\}\;,
\nonumber\\
H_F &=& \sum_{{\bf k}\lambda} \hbar \omega_k \hat{c}^{\dagger}_{{\bf k}
\lambda} \hat{c}_{{\bf k}\lambda} 
\;,\\
H_{AI} &=& - {\bf d} \sum_i \left( \sigma_i + \sigma_i^\dagger \right) 
{\bf E}_{in}({\bf r}_i,t)
\;,
\nonumber\\
H_{AF} &=& - \hbar \sum_i \sum_{{\bf k}\lambda} g_{{\bf k}\lambda}^* 
\hat{c}^{\dagger}_{{\bf k}\lambda} \exp\left( - i {\bf k} {\bf r} \right)
\left( \sigma_i + \sigma_i^\dagger \right) + H.c. 
\;,
\nonumber\\
&&g_{{\bf k}\lambda} = i \sqrt{\frac{2\pi\omega_k}{\hbar V}} {\bf d} 
{\bf e}_{{\bf k}\lambda} 
\;.\nonumber
\end{eqnarray}
$H_A$ is the 
Hamiltonian of the freely moving atoms, $H_F$ is the Hamiltonian of the 
free photons and $H_{AI}$ describes the interaction of the atoms with 
the incident field ${\bf E}_{in}$. The transition operator is 
$\sigma^\dagger = |2\rangle \langle1|$. The exchange of photons between 
the ground and excited states of the atoms is incorporated by the atom-field 
interaction $H_{AF}$. A rotating wave approximation has not been made. 
${\bf d}={\bf d}^*$ is the matrix element of the atomic dipole moment. 
$g_{{\bf k}\lambda}$ represents therefore the coupling strength of an 
atom to the photons. Transitions between the internal states of the atoms go 
back to the coupling of the atom to the incident field on one hand, and to 
the emission and absorption of photons on the other.

We assume that all contact interactions can be neglected. In order to find the conditions under which this approximation is valid, it is necessary to estimate the ratio $U_d/U_g$, where $U_d$ and $U_g$ are the mean energies of the dipole-dipole interaction and the ground-state collisions, respectively. Treating the ground-state collisions in terms of s-wave scattering and restricting to dipole optical transitions we get the following inequality\cite{WAL97}
\begin{equation}
\label{ratio}
\frac{U_d}{U_g} \gg
\frac{3}{8} \frac{s}{a_s k_a}
\;,
\end{equation}
where $s$ is the saturation parameter of the atomic transition, $a_s$ is the scattering length and $k_a = \omega_a/c$. The scattering length $a_s$ is a quantity of the order of $1\ nm$. In the optical domain $k_a$ is of the order of $0.01\ nm^{-1}$, and we get the following estimate
\begin{equation}
\label{estimate}
\frac{U_d}{U_g} \gg 37.5 s
\;.
\end{equation}
Thus, the ground-state collisions are negligible if the saturation parameter $s$ is of the order of $0.01$ or higher. This shows, that we don't need very high values of the parameter $s$ in order to neglect contact interactions.

To take into account quantum statistical effects for the two-level atoms, we 
turn to quantized matter fields represented by two-component vectors
\begin{equation}
\hat{\psi}({\bf r},t)=\hat{\psi}_1({\bf r},t)|1\rangle+
\hat{\psi}_2({\bf r},t)|2\rangle 
\;.\nonumber
\end{equation}
The respective bosonic equal time commutators are
\begin{eqnarray}\label{komm-equal}
\left[ \hat{\psi}_i({\bf r}) , \hat{\psi}_j({\bf r}') \right] &=& \left[ 
\hat{\psi}_i({\bf r}) , \hat{\psi}_j({\bf r}') \right] =0 
\;,\nonumber\\
\left[ \hat{\psi}_i({\bf r}) , \hat{\psi}_j^\dagger({\bf r}') \right] &=& 
\delta_{ij} \delta\left( {\bf r}-{\bf r}' \right) 
\;,\quad 
i,j=1,2. 
\end{eqnarray}
In second quantization the Hamiltonian operator (\ref{ham-many-part}) takes 
then the form:
\begin{eqnarray}\label{ham-second-quant}
{\cal H} &=& \sum_{j=1}^2 \int d{\bf r} \hat{\psi}_j^\dagger({\bf r}) 
\left( - \frac{\hbar^2 \nabla^2}{2m} \right) \hat{\psi}_j({\bf r}) + 
\int d{\bf r} \hat{\psi}_2^\dagger({\bf r}) \hbar \omega_a 
\hat{\psi}_2({\bf r}) + H_F\\ 
&-& \int d{\bf r} {\bf d} {\bf E}_{in}\hat{\psi}_1^\dagger({\bf r}) 
\hat{\psi}_2({\bf r}) - \hbar \int d{\bf r} \sum_{{\bf k}\lambda} 
g_{{\bf k}\lambda}^* \hat{c}^{\dagger}_{{\bf k}\lambda} \exp\left( 
- i {\bf k} {\bf r} \right)\left\{ \hat{\psi}_2^\dagger({\bf r}) 
\hat{\psi}_1({\bf r}) + \hat{\psi}_1^\dagger({\bf r}) 
\hat{\psi}_2({\bf r})\right\}- H.c.\nonumber
\end{eqnarray}
Making use of the Hamiltonian (\ref{ham-second-quant}) and the commutator 
relations we obtain the following Heisenberg equations of motion for the 
atomic and field operators:
\begin{mathletters}
\label{heis}
\begin{eqnarray} 
i \hbar \frac{\partial\hat{\psi}_1({\bf r},t)}{\partial t} &=& - 
\frac{\hbar^2 \nabla^2}{2m} \hat{\psi}_1({\bf r},t) - {\bf d} 
{\bf E}_{in}^-({\bf r},t) \hat{\psi}_2({\bf r},t)
\nonumber\\
&-& 
\hbar \sum_{{\bf k}\lambda} g_{{\bf k}\lambda}^* 
\hat{c}^{\dagger}_{{\bf k}\lambda}(t) \exp\left( - i {\bf k} {\bf r} \right) 
\hat{\psi}_2({\bf r},t) - \hbar \hat{\psi}_2({\bf r},t) \sum_{{\bf k}\lambda} 
g_{{\bf k}\lambda} \exp\left( i {\bf k} {\bf r} \right) 
\hat{c}_{{\bf k}\lambda}(t) 
\;,\label{heis-a}\\
i \hbar \frac{\partial\hat{\psi}_2({\bf r},t)}{\partial t} &=& - 
\frac{\hbar^2 \nabla^2}{2m} \hat{\psi}_2({\bf r},t) + \hbar \omega_a 
\hat{\psi}_2({\bf r},t) - {\bf d} {\bf E}_{in}^+({\bf r},t)
\hat{\psi}_1({\bf r},t)
\nonumber\\
&-& \hbar \sum_{{\bf k}\lambda} g_{{\bf k}\lambda}^* 
\hat{c}^{\dagger}_{{\bf k}\lambda}(t) \exp\left( - i {\bf k} {\bf r} 
\right) \hat{\psi}_1({\bf r},t) - \hbar \hat{\psi}_1({\bf r},t) 
\sum_{{\bf k}\lambda} g_{{\bf k}\lambda} \exp\left( i {\bf k} {\bf r} \right) 
\hat{c}_{{\bf k}\lambda}(t) 
\;,\label{heis-b}\\
i \hbar \frac{\partial \hat{c}_{{\bf k}\lambda}(t)}{\partial t} &=& \hbar 
\omega_k \hat{c}_{{\bf k}\lambda}(t) - \hbar g_{{\bf k}\lambda}^* 
\int d{\bf r} \exp\left( - i {\bf k} {\bf r} \right) \left[ 
\hat{\psi}_2^\dagger({\bf r},t) \hat{\psi}_1({\bf r},t) + 
\hat{\psi}_1^\dagger({\bf r},t) \hat{\psi}_2({\bf r},t) \right] 
\;,\label{heis-c}
\end{eqnarray}
\end{mathletters}
where ${\bf E}_{in}^\pm$ are the positive and negative frequency parts of the 
incident classical electric field.
\narrowtext

\section{The local-field approach}

From the physical point of view we are in fact investigating the propagation 
of an incident electric field through a particular medium. It is therefore 
reasonable to try to recover the fundamental quantities and relations of the 
well-known theory of electromagnetic fields in media~\cite{BOR68,JAC75eng}. 
There one regards the medium as being composed of interacting induced point 
dipoles in otherwise empty space. The incident field 
${\bf E}_{in} ({\bf r},t)$ causes the dipoles to emit radiation. The electric 
field to which a single dipole at the position $({\bf r},t)$ effectively 
reacts is then the resulting microscopic local field $\hat{{\bf E}}_{loc} 
({\bf r},t)$. On the other hand, the Maxwell equations in media refer to a 
macroscopic or mean electric field $\hat{{\bf E}}_{mac} ({\bf r},t)$ which 
is obtained by averaging over a region which contains a large number of 
atoms. $\hat{{\bf E}}_{mac} ({\bf r},t)$ and $\hat{{\bf E}}_{loc} ({\bf r},t)$ 
differ by the local-field correction or Lorentz-Lorenz correction. This 
correction is proportional to the macroscopic electric polarization 
$\hat{{\bf P}} ({\bf r},t)$ which also appears in the Maxwell equations. 
We will now show that the  dynamical operator equations which we obtain 
from (\ref{heis}) reflect in fact the physical structure of this {\it local-field approach}.

The formal solution of (\ref{heis-c}) for the photon operators is
\begin{eqnarray}\label{solution-photon}
\hat{c}_{{\bf k}\lambda}(t) 
&=& 
\Biggl\{ 
     \hat c_{{\bf k}\lambda}(0)
     \nonumber\\ 
     + i g_{{\bf k}\lambda}^* 
     \int_0^t 
   &dt'& 
     \int d{\bf r}' 
     \exp
     \left( 
          i \omega_k t' - i {\bf k} {\bf r}'
     \right) 
     \left[ 
          \hat \psi_2^\dagger({\bf r}',t') 
          \hat{\psi}_1({\bf r}',t') 
          + 
          \hat{\psi}_1^\dagger({\bf r}',t') 
          \hat{\psi}_2({\bf r}',t') 
     \right] 
\Biggr\} 
\exp 
\left( 
     - i \omega_k t 
\right)
\;, 
\end{eqnarray}
where the first term $\hat{c}_{{\bf k}\lambda}(0)$ refers to the free-space 
photon field and the second one goes back to the interaction with the atoms.

To study the back reaction of the photons on matter we insert 
(\ref{solution-photon}) in (\ref{heis-a}) and (\ref{heis-b}) and rearrange the matter operators with the help of the commutator relations (\ref{komm-equal}). For details of the calculations see Appendix A. Introducing the polarization operator $\hat{\bf P}^({\bf r},t)$ of the ultracold ensemble
\begin{equation}
\label{P}
\hat{\bf P}^({\bf r},t)
=
{\bf d}
\hat\psi_1^\dagger({\bf r},t)
\hat\psi_2({\bf r},t)
+ 
H.c.
=
\hat{\bf P}^+({\bf r},t)
+
\hat{\bf P}^-({\bf r},t)
\;,
\end{equation}
we obtain as a central result the dynamical equations for the operators of the two matter fields
\begin{mathletters}
\label{matter}
\begin{eqnarray} 
&i& \hbar \frac{\partial\hat{\psi}_1({\bf r},t)}{\partial t} 
=
- 
\frac{\hbar^2 \nabla^2}{2m} \hat{\psi}_1({\bf r},t) - {\bf d} 
\hat{{\bf E}}_{loc}^-({\bf r},t) \hat{\psi}_2({\bf r},t)
\nonumber\\
&-& \hbar 
\sum_{{\bf k}\lambda} 
g_{{\bf k}\lambda}^* 
\hat{c}^{\dagger}_{{\bf k}\lambda}(0) 
\exp\left( - i {\bf k} {\bf r} + i \omega_k t \right) 
\hat{\psi}_2({\bf r},t)
\nonumber\\
&-& \hbar 
\hat{\psi}_2 ({\bf r},t) 
\sum_{{\bf k}\lambda} g_{{\bf k}\lambda}
\hat{c}_{{\bf k}\lambda}(0) 
\exp\left( i {\bf k} {\bf r} - i \omega_k t \right) 
\;, 
\label{matter-a}\\
i \hbar \frac{\partial\hat{\psi}_2({\bf r},t)}{\partial t} 
&=& 
- 
\frac{\hbar^2 \nabla^2}{2m} \hat{\psi}_2({\bf r},t) 
+ 
\hbar 
\left(
     \omega_a + \delta - i \gamma/2
\right) 
\hat\psi_2({\bf r},t) 
- 
{\bf d} \hat{\bf E}_{loc}^+({\bf r},t) 
\hat\psi_1({\bf r},t)
\nonumber\\
&-& 
\hbar 
\sum_{{\bf k}\lambda} 
g_{{\bf k}\lambda}^* 
\hat{c}^{\dagger}_{{\bf k}\lambda}(0) 
\exp
\left( - i {\bf k} {\bf r} + i \omega_k t \right) 
\hat\psi_1({\bf r},t) 
\nonumber\\
&-& 
\hbar 
\hat{\psi}_1({\bf r},t) 
\sum_{{\bf k}\lambda} 
g_{{\bf k}\lambda} 
\hat{c}_{{\bf k}\lambda}(0)
\exp
\left( i {\bf k} {\bf r} - i \omega_k t \right) 
\;, \label{matter-b}
\end{eqnarray}
\end{mathletters}
where we have introduced
\begin{equation}
\label{def-e-local}
\hat{{\bf E}}_{loc}^\pm({\bf r},t) = {\bf E}_{in}^\pm({\bf r},t) + \int 
d{\bf r}' 
\nabla \times \nabla \times 
\frac { \hat{\bf P}^\pm \left( {\bf r}',t-R/c \right)}{R}
\;,
\end{equation}
and $\nabla\times$ refers to the point ${\bf r}$.

Eq. (\ref{def-e-local}) shows that $\hat{{\bf E}}_{loc}^\pm({\bf r},t)$ is the superposition of the incident field ${\bf E}_{in}^\pm({\bf r},t)$ and the electric field which is obtained by adding up at the space-time point $({\bf r},t)$ the electric dipole radiation of all other atoms. $\hat{{\bf E}}_{loc}^\pm$ is therefore the operator of the 
{\it local field}. It is correct for all distances including the near-field zone. As one should expect it is this local field which 
drives the inner atomic transition in (\ref{matter}). Because of this 
fact the matter field equations (\ref{matter}) together with 
(\ref{def-e-local}) and (\ref{P}) are nonlinear. This follows directly without approximation from the Heisenberg equations (\ref{heis}). The physical reason is that the electromagnetic field which propagates through the medium causes the atoms to emit and absorb photons of the dipole radiation. 

Note that in the eq.(\ref{def-e-local}) a small volume around the observation point ${\bf r}$ is excluded from the integration. The eq.(\ref{def-e-local}) can be written down in another form, where the integration is carried out over the whole space, besides there is an additional term in the r.h.s. which containes a $\delta$-function\cite{MOR95,JAC75eng}. These two forms of integro-differential equation are totally equivalent and give the same results.

The equations (\ref{matter}) are nothing but the atom-optical analogue of the usual optical Bloch equations\cite{ALL78,BOW93}. They describe the dynamical evolution of the second quantized matter in the field of electromagnetic radiation. In the Appendix B we shall show that under certain conditions the optical Bloch equations which describe the dynamical evolution of the first quantized matter can be derived from the equations (\ref{matter}).

Treating the quantum gas as a medium, it is important to establish a 
connection to the observables of the quantum electrodynamics in media. 
Having an application to an ultracold gas in mind, we have to relate 
the local field $\hat{{\bf E}}_{loc}({\bf r},t)$ to the {\it macroscopic 
(or mean) field} $\hat{{\bf E}}_{mac}({\bf r},t)$ which is obtained by 
averaging in space over a region which contains a great number of atoms. For a very low density ultracold Bose gas the region on which the macroscopic field is averaged may contain actually fewer than one atom, yet the averaging technique should still be valid due to atomic delocalization.

We generalize the idea of Lorentz as it is described in\cite{BOW93,BOR68,JAC75eng} and assume the local field to be a sum of several contributions:
\begin{equation}
\label{L-method}
\hat {\bf E}_{loc}({\bf r},t) = 
\hat {\bf E}_{near}({\bf r},t) + 
\hat {\bf E}_{mac}({\bf r},t) -
\hat {\bf E}_P({\bf r},t)
\;.
\end{equation}
Let's consider a macroscopically small, but microscopically large volume ${\cal V}$ around ${\bf r}$. $\hat {\bf E}_{near}$ is a contribution from the dipoles within ${\cal V}$. $\hat {\bf E}_{mac}$ is the macroscopic field obtained from the field of all dipoles in the medium. Because the dipoles in ${\cal V}$ are already taken into account in $\hat {\bf E}_{near}$, we have to subtract from $\hat {\bf E}_{mac}$ the contribution $\hat {\bf E}_P$ of these dipoles as it is obtained in the averaged continuum approximation. $\hat {\bf E}_P$ is therefore a macroscopic quantity. It can be related to the polarization, which is an averaged quantity too, according to
\begin{equation}
\label{EP}
\hat {\bf E}_P({\bf r},t) = - \frac{4\pi}{3} \hat {\bf P}({\bf r},t)
\;.
\end{equation}
For a novel derivation see\cite{BOW93}. In the following we restrict to the purely classical local-field correction which corresponds to a vanishing $\hat {\bf E}_{near}$ (comp.\cite{BOW93,BOR68,JAC75eng}). Thus, we see that, because of the influence of the near dipole-dipole interactions, the local field is obtained from the macroscopic field in adding the {\it local-field correction} (comp. e.g.~\cite{BOW93,BOR68,JAC75eng}).
\begin{equation}
\label{local-corr}
\hat{{\bf E}}_{loc}^\pm({\bf r},t) = 
\hat{{\bf E}}_{mac}^\pm({\bf r},t) + 
\frac{4\pi}{3}\hat{{\bf P}}^\pm({\bf r},t) 
\;. 
\end{equation}
This equation is also often called the {\it Lorentz-Lorenz relation}. The 
propagation of the operator of the macroscopic field is thereby given 
by Maxwell wave equation for a charge-free and current-free polarization medium
\begin{equation}
\label{maxwell-equ}
\nabla \times \nabla \times 
\hat{{\bf E}}_{mac}^\pm({\bf r},t) 
= 
- \frac{1}{c^2} 
\frac{\partial^2 \hat{{\bf E}}_{mac}^\pm({\bf r},t)}{\partial t^2} 
-
\frac{4\pi}{c^2} 
\frac{\partial^2 \hat{{\bf P}}^\pm({\bf r},t)}{\partial t^2} 
\;. 
\end{equation}
When taking the expectation value, the operator $\hat{{\bf P}}({\bf r},t)$ 
leads to the averaged dipole moment per unit volume.

To sum up: Taking together eqs.~(\ref{maxwell-equ}) and the 
truncated eq.~(\ref{matter}) with (\ref{local-corr}), we have obtained a self-consistent system of coupled differential equations for the collective dynamics of the fields 
$\hat{{\bf E}}_{mac}({\bf r},t)$ and $\hat{\psi}_{1,2}({\bf r},t)$. Note 
that when dealing with the Maxwell propagation equation (\ref{maxwell-equ}), 
the existence of the incident field ${\bf E}_{in}$ has to be taken into 
account. It initiates the polarization of the medium. With (\ref{local-corr}) 
and (\ref{P}) the matter field equations (\ref{matter}) are 
nonlinear. This is the direct consequence of the fact that matter couples 
necessarily to the local field as the driving field for internal atomic 
transitions and not to the macroscopic electric field.

\section{Nonlinear matter equation and optical properties}

\subsection{General case}

In the next step we decouple the dynamical equations for the two matter field 
operators. We substitute (\ref{local-corr}) in (\ref{matter-a}) and 
(\ref{matter-b}) and pass to the reference frame rotating with the frequency 
$\omega_L$ of the incident field which is assumed to be monochromatic

\begin{eqnarray}\label{transform-rot} 
\hat{{\bf E}}_{mac}^+({\bf r},t) &=& \hat{\bf \cal E}_{mac}^+({\bf r}) \exp 
\left( - i \omega_L t \right)\;,\nonumber\\ 
\hat{\psi}_2({\bf r},t) &=& \hat{\phi}_2({\bf r},t)\exp \left( - i \omega_L t 
\right)\;,
\end{eqnarray}
to obtain
\begin{mathletters}
\label{nonlinear}
\begin{eqnarray}
i \hbar \frac{\partial \hat \psi_1}{\partial t} 
&=& 
- \frac{\hbar^2 \nabla^2}{2m} \hat \psi_1 
- \frac{\hbar}{2} \hat{\Omega}^-({\bf r}) \hat \phi_2 
- \frac{4\pi}{3} d^2 \hat \phi_2^\dagger \hat \phi_2 \hat \psi_1
+ \hat G_1
\;, 
\label{nonlinear-a}\\
i \hbar \frac{\partial\hat{\phi}_2}{\partial t} &=& - 
\frac{\hbar^2 \nabla^2}{2m}\hat{\phi}_2 - \frac{\hbar}{2} 
\hat{\Omega}^+({\bf r}) \hat{\psi}_1 - \frac{4\pi}{3} d^2 
\hat{\psi}_1^\dagger\hat{\psi}_1 \hat{\phi}_2\nonumber\\
&\quad&
- \hbar \left( \Delta + i \gamma/2 \right) \hat{\phi}_2 
+ \hat G_2
\;, 
\label{nonlinear-b}
\end{eqnarray}
\end{mathletters}
with the detuning $\Delta=\omega_L-\omega_a-\delta$. The position dependent 
Rabi frequency 
$\hat{\Omega}^\pm({\bf r})=2{\bf d}\hat{\bf \cal E}_{mac}^\pm({\bf r})/\hbar$ is related to the macroscopic electric field. The noise terms in the eqs.(\ref{nonlinear}) are
\begin{eqnarray}
\label{G}
\hat G_1
\left(
     {\bf r},t
\right)
&=&
- \hbar
\left[
     \hat \Gamma_1^\dagger
     \left(
          {\bf r},t
     \right)
     \hat \phi_2
     + \hat \phi_2
     \hat \Gamma_2
     \left(
          {\bf r},t
     \right)
\right]
\;,
\nonumber\\
\hat G_2
\left(
     {\bf r},t
\right)
&=&
- \hbar
\left[
     \hat \Gamma_2^\dagger
     \left(
          {\bf r},t
     \right)
     \hat \psi_1
     + \hat \psi_1
     \hat \Gamma_1
     \left(
          {\bf r},t
     \right)
\right]
\;,
\end{eqnarray}
with operators
\begin{eqnarray}
\label{Gamma}
\hat \Gamma_1
\left(
     {\bf r},t
\right)
&=&
\sum_{{\bf k}\lambda}
g_{{\bf k}\lambda}
\hat c_{{\bf k}\lambda}(0)
\exp
\left[
    - i
    \left(
         \omega_k - \omega_L
    \right) 
    t
    + i {\bf k} {\bf r}
\right]
\;,
\nonumber\\
\hat \Gamma_2
\left(
     {\bf r},t
\right)
&=&
\sum_{{\bf k}\lambda}
g_{{\bf k}\lambda}
\hat c_{{\bf k}\lambda}(0)
\exp
\left[
    - i
    \left(
         \omega_k + \omega_L
    \right) 
    t
    + i {\bf k} {\bf r}
\right]
\;,
\end{eqnarray}
which give the effect of vacuum fluctuations on the atomic quantum field.

We assume sufficiently large detuning such that the spontaneous emission gives a small contribution. We may therefore apply {\it the adiabatic approximation} to (\ref{nonlinear-b}) (comp.~\cite{ZHA94a})
\begin{equation}\label{adiabatic-sol}
\hat{\phi}_2({\bf r},t) = 
- 
\frac
{\hat{\Omega}^+({\bf r}) \hat{\psi}_1({\bf r},t)}
{ 2\left( \hat{\Delta}_l({\bf r},t) + i \gamma/2 \right) }
+
\frac
{ \hat G_2 ({\bf r},t) }
{ \hbar \left( \hat{\Delta}_l({\bf r},t) + i \gamma/2 \right) }
\;, 
\end{equation}
and eliminate the field $\hat{\psi}_2$ from our scheme. The position 
dependent {\it local detuning} introduced in (\ref{adiabatic-sol})
\begin{equation}
\label{def-local-det}
\hat{\Delta}_l({\bf r},t) = 
\Delta + 
\frac{4\pi}{3\hbar} d^2 
\hat{\psi}_1^\dagger({\bf r},t) 
\hat{\psi}_1({\bf r},t)  
\end{equation}
is an operator depending on the density operator of the ground state. We 
interpret this as a shift of the internal energy levels which may increase or decrease with increasing density depending on the sign of the detuning $\Delta$. Corresponding frequency shifts can be found in nonlinear optics where they lead to the effects mentioned in the introduction. Substituting (\ref{def-local-det}) in 
(\ref{nonlinear-a}) we obtain as the intended result the nonlinear dynamical 
equation for the matter field operator $\hat{\psi}_1({\bf r},t)$
\begin{eqnarray}
\label{nonlinear-equation} 
i\hbar\frac{\partial \hat{\psi}_1({\bf r},t)}{\partial t}
&=& 
\left\{ 
-\frac
{\hbar^2\nabla^2}{2m} +\frac{\hbar}{4} \frac{\Delta-i\gamma/2}{
\hat{\Delta}_l^2({\bf r},t)+\gamma^2/4} \left| \hat{\Omega}^+({\bf r})
\right|^2 
+ \hat V_R
\right\} 
\hat{\psi}_1({\bf r},t)
\;,
\nonumber\\
\hat V_R
&=&
\hbar
\frac{\Delta-i\gamma/2}{\hat{\Delta}_l^2({\bf r},t)+\gamma^2/4}
\left[
     \frac{1}{2}
     \left(
          \hat \Gamma_2 + \hat \Gamma_1^\dagger
     \right)
     \hat \Omega^+
     +
     \frac{1}{2}
     \left(
          \hat \Gamma_2^\dagger + \hat \Gamma_1
     \right)
     \hat \Omega^-
     +
     \left|
          \hat \Gamma_1^\dagger + \hat \Gamma_2
     \right|^2
\right]
\;.  
\end{eqnarray}
It couples to the macroscopic electric field via the Rabi frequency 
$\hat{\Omega}^\pm({\bf r})$. Here $\hat V_R$ is a random potential caused by vacuum fluctuations. Its vacuum average $\langle \hat V_R \rangle$ vanishes.

On the assumptions stated in Sect.~I, the equation (\ref{nonlinear-equation}) 
is general. No approximation apart from the adiabatic approximation has been made. The dipole-dipole interaction mediated through the exchange of dynamically produces photons is completely contained. 
The influence of the atoms is taken into account in the Maxwell equations 
(\ref{maxwell-equ}). The local-field correction is included. It is not 
assumed that the density is low. The nonlinearity goes back to the density 
dependent local detuning $\hat{\Delta}_l({\bf r},t)$ of 
eq.~(\ref{def-local-det}). For increasing density and positive detuning $\Delta$ the local detuning grows and correspondingly the nonlinear term in (\ref{nonlinear-equation}) representing the coupling to the macroscopic electric field becomes smaller. On the other hand, for negative detuning the absolute value of the local detuning decreases and the nonlinearity becomes greater. When (\ref{nonlinear-equation}) and the Maxwell wave equation (\ref{maxwell-equ}) are solved, all possible effects on the center-of-mass motion of the atoms can be worked out.

Because we are mainly interested in atom optical problems and want to study the coherent evolution of the center-of-mass motion of the gas, we shall neglect spontaneous emission. This is valid for situations where the absolute value of the local detuning is much bigger than the spontaneous emission rate $\left|\Delta_l\right| \gg \gamma$. In order to do this we drop in the following the random potential $\hat V_R$ and the spontaneous emission rate $\gamma$ from our equations.

\subsection{Special cases}

The equations which are usually used for the description of an ultracold 
ensemble put in an external laser field may be derived from 
(\ref{nonlinear-equation}) as {\it limiting cases}. Neglecting dipole-dipole interactions we have 
$\hat{\Delta}_l = \Delta$ and $\hat{{\bf E}}_{loc}^\pm = \hat{{\bf 
E}}_{mac}^\pm = {\bf E}_{in}^\pm$. The eq.(\ref{nonlinear-equation}) reduces 
to the well-known equation of the single particle theory~\cite{CHE85} 
\begin{equation}
\label{single-part}
i\hbar\frac{\partial \hat{\psi}_1({\bf r},t)}{\partial t}= 
\left\{ 
-\frac{\hbar^2\nabla^2}{2m} +\frac{\hbar}{4} \frac{\left| \Omega_{in}^+
\right|^2}{\Delta} 
\right\} 
\hat{\psi}_1({\bf r},t)
\;.
\end{equation}

Expanding (\ref{nonlinear-equation}) to the lowest order in the density 
$\hat{\psi}_1^\dagger \hat{\psi}_1$ we get an equation of the Gross-Pitaevskii type which is here obtained in the local-field approach:
\begin{equation}\label{gross-pia}
i\hbar\frac{\partial \hat{\psi}_1({\bf r},t)}{\partial t}
= 
\left\{
     -\frac{\hbar^2\nabla^2}{2m} 
     +
     \frac{\left| \hat{\Omega}^+({\bf r}) \right|^2}{\Delta}
     \left[
          \frac{\hbar}{4}-
          \frac{2\pi}{3} \frac{d^2}{\Delta}
          \hat{\psi}_1^\dagger({\bf r},t) \hat{\psi}_1({\bf r},t)
     \right] 
\right\} 
\hat{\psi}_1({\bf r},t)
\;. 
\end{equation}
It may be applied fow low densities. A similar equation has been obtained in~\cite{ZHA94a}.

Another approximative approach is the following: Introducing the standard 
(linear) atomic polarizability $\alpha=-d^2/\hbar\Delta$, 
we can rewrite the nonlinear term in eq.(\ref{nonlinear-equation}) as
\begin{equation}\label{pre-wallis}
\frac{\hbar}{4} 
\frac{\left| \hat{\Omega}^+({\bf r})\right|^2}
{
\Delta
\left[
     1 - 
     \frac{8\pi}{3}\alpha
     \hat\psi_1^\dagger\hat\psi_1
     +
     \left(
           \frac{4\pi}{3}
     \right)^2
     \alpha^2
     \hat\psi_1^\dagger \hat\psi_1 \hat\psi_1^\dagger \hat\psi_1
\right]
}
\;.
\end{equation}
Neglecting the term in the denominator proportional to the density squared, we get
\begin{equation}\label{wallis}
i\hbar\frac{\partial \hat{\psi}_1({\bf r},t)}{\partial t}
= 
\left\{
      -\frac{\hbar^2\nabla^2}{2m} 
      +
      \frac{\hbar}{4} 
      \frac
      {\left| \hat{\Omega}^+({\bf r}) \right|^2} 
      {
       \Delta
       \left(
            1 - 
            \frac{8\pi}{3} \alpha \hat\psi_1^\dagger \hat\psi_1
       \right)
      }
\right\} 
\hat{\psi}_1({\bf r},t)
\;,
\end{equation}
which agrees with the eq.(3.12) of the paper~\cite{WAL97} for vanishing 
contact interaction and vanishing gravitational potential apart from a factor $2$ (we have $8\pi/3$ instead of $4\pi/3$).

\subsection{Optical properties}

We return to the general case of Sect.~IV to discuss the optical properties 
of our ultracold Bose gas. We insert the adiabatic solution 
(\ref{adiabatic-sol}) into the polarization (\ref{P}) to obtain
\begin{equation}
\label{media-equ}
\hat{{\bf P}}^+({\bf r},t) = \hat{\chi}({\bf r},t) \hat{{\bf E}}_{mac}^+
({\bf r},t)
\end{equation}
and $\hat{{\bf P}}^-$ correspondingly with the operator of dielectric 
susceptibility of our medium
\begin{equation}
\label{def-suszept}
\hat{\chi}({\bf r},t)=
\frac{\alpha\hat{\psi}_1^\dagger({\bf r},t) 
\hat{\psi}_1({\bf r},t)}{1-\frac{4\pi}{3}\alpha\hat{\psi}_1^\dagger({\bf 
r},t) 
\hat{\psi}_1({\bf r},t)}
\;,
\end{equation}
where $\alpha$ is the atomic polarizability stated above. (\ref{media-equ}) 
with (\ref{def-suszept}) corresponds the well-known Lorentz-Lorenz formula. 
Inserting (\ref{media-equ}) in (\ref{maxwell-equ}) we find together with 
(\ref{nonlinear-equation}) the new system of coupled equations for the 
quantum fields $\hat{{\bf E}}_{mac}^\pm({\bf r},t)$ and $\hat{\psi}_1({\bf 
r},t)$ as it is valid in the adiabatic approximation.

Assuming for this system that the spatial variations of the atomic density 
are not very large so that 
$div\, \hat{\bf E}_{mac}^\pm \approx 0$, we have
\begin{equation}
\label{maxwell-media}
\nabla^2 \hat{\bf \cal E}_{mac}^\pm  + 
k_L^2 \hat{n}^2 \hat{\bf \cal E}_{mac}^\pm = 0
\;,
\end{equation}
whereby the refractive index satisfies the Clausius-Mossotti relation:

\begin{equation}\label{clausius}
\hat{n}^2 =\frac{1 + \frac{8\pi}{3} \alpha \hat{\psi}_1^\dagger \hat{\psi}_1
}{1 - \frac{4\pi}{3} \alpha \hat{\psi}_1^\dagger \hat{\psi}_1}
\end{equation}
Based on our system of coupled equations and the related optical 
interpretation, we turn now to an application and discuss an atom optical 
device for ultracold atoms, which can be used as a simple model for a beam 
splitter.

\section{Diffraction of an ultracold atomic beam from a strong standing 
light wave}

We have to solve both the equation of the quantum field 
(\ref{nonlinear-equation}) and Maxwell's equation (\ref{maxwell-media}) 
with (\ref{clausius}). In general this has to be done in a self-consistent 
way and it will be difficult to do so analytically. Nevertheless, some 
simplifying approximations can be made when applying the whole formalism to 
the diffraction of an ultracold atomic beam by a standing electromagnetic 
wave. For this we generalize and modify the approach of~\cite{ZHA94a}.

We consider the following setup: An incident atomic beam moves in 
$z$-direction, perpendicular to a standing laser beam consisting of two 
counter propagating waves along the $y$-axis with wave vectors $+n{\bf k}_L$ 
and $-n{\bf k}_L$, respectively, and with a Gaussian envelope. The motion 
in $z$-direction remains classical during the whole evolution and only the 
diffraction in $y$-direction is treated quantum mechanically. In order to 
get a distinct diffraction pattern, we assume that the width of the atomic 
wave packet $w_y$ is sufficiently large. Additionally the variations of the 
density should be small on a scale of the effective wavelength of the laser 
($w_y \gg 2\pi/nk_L$). In this case the atoms can be described as a 
homogeneous medium. The effect of the atoms on the laser beam is purely 
dispersive and only the wavelength will be shifted. The solution to 
(\ref{maxwell-media}) with (\ref{clausius}) is then given by
\begin{equation}\label{sol-maxwell}
\left| \Omega^+ \right|^2 = \left| \Omega_0 \right|^2 \exp\left( - z^2 / 
w_L^2 \right) \cos^2 nk_L y \;. 
\end{equation}

To solve the equation of motion for the atomic quantum field, we assume 
that we are in the Raman-Nath regime and that in equation 
(\ref{nonlinear-equation}) the kinetic term can be neglected during the 
interaction of the atoms with the electromagnetic field. This approximation 
is valid for heavy atoms or if the interaction is so strong that atoms can 
take up momentum without changing considerably the velocity. In this case 
the density of the atoms remains unaltered and their phase changes. With 
the definition 
\begin{equation}\label{eff-volume}
V_0 = \frac{4\pi}{3\hbar} \frac{d^2}{\Delta} 
\;, 
\end{equation}
of the characteristic volume, we then get for the equation of motion of the 
atomic beam
\begin{equation}\label{equ-motion-diffract}
i \frac{\partial \hat{\psi}_1}{\partial t} = \frac{1}{4 \Delta \left( 1 + 
V_0 \rho_1 \right)^2} \left| \Omega^+ \right|^2 \hat{\psi}_1 
\;, 
\end{equation}
where $\rho_1=\hat{\psi}_1^\dagger \hat{\psi}_1$ is the density of atoms in 
the ground state. This also guarantees that the conditions leading to 
(\ref{sol-maxwell}) are always satisfied, because the density distribution 
remains constant in time and only the phase of the atomic quantum field 
changes. Making use of the assumption that the motion in $z$-direction is 
purely classical, we change the variable $t=z/v_g$ in 
(\ref{equ-motion-diffract}) with $v_g$ being the group velocity of the atomic 
beam. Then the solution of eq.(\ref{equ-motion-diffract}) for $z \gg w_L$ 
(in the far zone) may be written in the following form 
\begin{equation}\label{sol-diffract-1}
\hat{\psi}_1(y,\infty) = \hat{\psi}_1(y,-\infty) \exp \left(
\int_{-\infty}^\infty \frac{-i\left| \Omega^+ \right|^2}{4\Delta 
v_g\left( 1 + V_0 \rho_1\right)^2} dz \right).
\end{equation}

Representing $\rho_1$ as a Gaussian wave packet with width $w_y$
\begin{equation}\label{density-profile}
\rho_1 = \rho_0 \exp\left( - y^2 / w_y^2 \right) 
\;,
\end{equation}
we substitute (\ref{sol-maxwell}) and (\ref{density-profile}) into 
(\ref{sol-diffract-1}) and get after integration
\begin{equation}\label{sol-diffract-2}
\hat{\psi}_1(y,\infty) = \hat{\psi}_1(y,-\infty) \exp \left(\frac{-i 4 g_0 
\cos^2 nk_L y}{\left[ 1 + V_0 \rho_0 \exp \left( - y^2 / w_y^2  \right)
\right]^2} \right), 
\end{equation}
where $g_0 = \frac{\Omega_0^2}{16\Delta} \frac{w_L}{v_g} \sqrt{\pi}$. 

The relative broadness of the wave packet compared to the laser wavelength 
can be expressed as $w_y \gg 2\pi/ nk_L$. Herewith we can represent the 
solution (\ref{sol-diffract-2}) in the form of a Fourier series expansion: 

\begin{equation}\label{expansion}
\hat{\psi}_1(y,\infty) = \hat{\psi}_1 \left(y,-\infty\right) e^{- i \tau} 
\sum_{q=-\infty}^\infty e^{i 2 q nk_L y } (-i)^q J_q(\tau) 
\;, 
\end{equation}
where $\tau = 2 g_0 / \left( 1 + V_0 \rho_0 \right)^2$ and $J_q$ 
is the $q$-th order Bessel function. Accordingly the probability $P_q$ to find 
the wave packet in a momentum state $q n k_L$ is given by
\begin{equation}\label{probabilities}
P_q = J_q^2(\tau)
\;,\quad 
q=-\infty, ..., + \infty 
\;,  
\end{equation}
with $P_0$ corresponding to the intensity of the outgoing beam which 
propagates in the same direction as the incident atomic beam. The angle of 
diffraction $\alpha_q$ for a particular momentum state is thereby given by
\begin{equation}\label{angles}
\tan\alpha_q = \frac{q n \hbar k_L}{m v_g}
\;. 
\end{equation}
Therefore the diffraction pattern as it follows from (\ref{probabilities}) 
depends on the initial density $\rho_0$ of the beam which is contained in 
$\tau$. If $\rho_0 = 0$, the formula (\ref{probabilities}) coincides with 
the formula (14) of paper~\cite{GRI85}.

The result (\ref{probabilities}) provides the possibility to obtain 
semi-quantitative estimates of the influence of the initial density $\rho_0$ 
of the ultracold atomic beam on the intensities of the various diffraction 
orders. Let's consider first the case of positive detuning $\Delta$. It is obvious that in this case if $\rho_0$ increases, the dimensionless effective interaction time $\tau$ will decrease. In addition, taking into account the form of the Bessel functions, the relative intensities of the diffraction orders with $q \not= 0$ decrease and $P_0$ increases. In the case of negative detuning $\Delta$ the absolute value of $\tau$ decreases with the increase of the density $\rho_0$. This implies that the effect of increasing density in the case of negative detuning $\Delta$ is to increase the coupling to the laser, enhancing the diffraction pattern. Physically these features originate from a 
shift of the internal energies of the atoms going back to the interaction 
with the other nearby atoms, as it is clearly expressed by equation 
(\ref{def-local-det}). The detuning of the laser beam changes locally with 
the density. With increasing density the coupling to the laser in equation 
(\ref{nonlinear-equation}) diminishes or increases depending on the detuning $\Delta$. Thus, our estimates show that with the increase of initial density, the atomic beam wave properties being manifested in the diffraction can be suppressed or enhanced due to dynamical dipole-dipole interactions. These effects are significant when $V_0 \rho_0 \sim 1$, i.e. when 
$\rho_0 \sim (\Delta/\gamma) k_L^3/\pi$. In the optical domain $k_L^3$ is of the order of $10^{15}\ cm^{-3}$. Taking into account that due to the fact that in our analysis $|\Delta| \gg \gamma$ the estimated value of the density required for large nonlinear effects at present is clearly larger than what can be obtained for alkali atoms in a magnetic trap.

\section{conclusion}

We have treated the interaction of a dense Bose gas with light in the local-field approach. As an application we have shown that it is possible to obtain analytical results for the the problem of scattering of an ultracold gas 
from a strong standing laser wave in the Raman-Nath regime. These results 
are not limited to low densities and include the dipole-dipole interaction 
consistently. Two effects had to be considered: First, the dynamical 
dipole-dipole interaction of nearby atoms shifts the atomic transition 
frequency. This gives a {\em locally} defined detuning between the atomic 
transition and laser frequency, which depends on the {\em local} density. 
Secondly, the atomic gas acts as a {\em medium} for the electromagnetic 
field and has to be taken into account in the {\em Maxwell equations}. 
While retardation of the electromagnetic field can always be neglected for 
short distances and gives only a negligible contribution to the level shift, 
retardation plays an important role at higher densities, and is necessary 
to describe the dispersion of electromagnetic waves.

Both effects give a contribution to the center-of-mass motion of the atoms 
in the gas and have to be considered simultaneously if the formalism is 
applied to atomic scattering. If variations of the density of the gas are 
only on length scales which are long compared to the laser frequency, the 
refractive index of the gas can be defined. The corresponding expression 
coincides with the well-known Clausius-Mossotti relation, and allows oneself 
to find a solution to the Maxwell equation alone, provided the conditions 
for the validity of the solution are preserved during the scattering process. 
This is guaranteed in the Raman-Nath regime, where the motion of the atoms 
is neglected and therefore the density remains constant. The result is again 
a standing wave, but the wavelength has changed according to the refractive 
index. With this, the recoil on the atoms is different compared to the case 
of no medium considered. Additionally, because of the density dependence of 
the atom-laser detuning, the time scale of scattering may be prolonged or decreased for high density gases, giving a suppression or enhancement of the scattering.

\section*{Acknowledgments}

This work has been supported by the Deutsche For\-schungsgemeinschaft 
and the Optikzentrum Konstanz. One of us (K.V.K.) would like to 
thank the members of the AG Audretsch at the University of Kon\-stanz 
for many interesting discussions and kind hospitality.

\section*{Appendix A}

To derive the eqs. (\ref{matter}) and (\ref{def-e-local}) which are a central result of the local-field approach, we insert the solution (\ref{solution-photon}) for photon operators into the eqs.(\ref{heis}) for the matter field operators. From (\ref{heis-a}) we obtain for the field operator of the lower state
\begin{eqnarray}
\label{heis-aa}
i &\hbar& \frac{\partial \hat\psi_1({\bf r},t)}{\partial t} 
=
- \frac{\hbar^2 \nabla^2}{2m} \hat\psi_1({\bf r},t) 
- {\bf d} {\bf E}_{in}^-({\bf r},t) \hat{\psi}_2({\bf r},t)
\nonumber\\ 
- &\hbar& \sum_{{\bf k}\lambda} g_{{\bf k}\lambda}^* 
\hat c^{\dagger}_{{\bf k}\lambda}(0) 
\exp
\left( 
    - i {\bf k} {\bf r} + i \omega_k t
\right) 
\hat\psi_2({\bf r},t) 
- 
\hbar \hat\psi_2({\bf r},t) 
\sum_{{\bf k}\lambda} g_{{\bf k}\lambda} 
\exp
\left( 
     i {\bf k} {\bf r} - i \omega_k t
\right) 
\hat c_{{\bf k}\lambda}(0)
\nonumber\\ 
&& 
+
\hat X ({\bf r},t)
+
\hat Y ({\bf r},t)
+
\hat Z ({\bf r},t)
\;,
\end{eqnarray}
where
\begin{eqnarray}
\label{operator-X}
\hat X ({\bf r},t) 
&=&
- i \hbar \int d{\bf r}' \int_0^t dt'
\hat\psi_1^\dagger({\bf r}',t')
\hat\psi_2({\bf r}',t')
\sum_{{\bf k}\lambda} 
\left|
      g_{{\bf k}\lambda}
\right|^2
\left\{
    \exp
    \left[
         i {\bf k}
         \left(
              {\bf r} - {\bf r}'
         \right)
         - i \omega_k
         \left(
              t - t'
         \right)
    \right]
    - c.c.
\right\}
\nonumber\\
& & \times \hat\psi_2({\bf r},t)
\;,
\end{eqnarray}
\begin{equation}
\label{operator-Y}
\hat Y ({\bf r},t)
=
-
i \hbar \int d{\bf r}' \int_0^t dt'
\sum_{{\bf k}\lambda} 
\left|
      g_{{\bf k}\lambda}
\right|^2
    \exp
    \left[
         i {\bf k}
         \left(
              {\bf r} - {\bf r}'
         \right)
         - i \omega_k
         \left(
              t - t'
         \right)
    \right]
\hat\psi_2({\bf r},t)
\hat\psi_2^\dagger({\bf r}',t')
\hat\psi_1({\bf r}',t')
\;,
\end{equation}
\begin{equation}
\label{operator-Z}
\hat Z ({\bf r},t) 
=
i \hbar \int d{\bf r}' \int_0^t dt'
\sum_{{\bf k}\lambda} 
\left|
      g_{{\bf k}\lambda}
\right|^2
    \exp
    \left[
         - i {\bf k}
         \left(
              {\bf r} - {\bf r}'
         \right)
         + i \omega_k
         \left(
              t - t'
         \right)
    \right]
\hat\psi_2^\dagger({\bf r}',t')
\hat\psi_1({\bf r}',t')
\hat\psi_2({\bf r},t)
\;.
\end{equation}

We turn to the evaluation of $\hat X({\bf r},t)$ first. With
\begin{equation}
\label{gkl}
g_{{\bf k}\lambda}
=
i 
\sqrt{\frac{2\pi\omega_k}{\hbar V}}
{\bf d} {\bf e}_{{\bf k}\lambda}
\end{equation}
and
\begin{equation}
\label{de2}
\left(
     {\bf d} {\bf e}_{{\bf k}\lambda}
\right)^2
=
\sum_{m,n=1}^3
d_m e_{{\bf k}\lambda}^m
d_n e_{{\bf k}\lambda}^n
\end{equation}
we find
\begin{eqnarray}
\label{Xsmn}
\hat X ({\bf r},t) 
&=&
-
\frac{i}{\hbar} 
\int d{\bf r}' \int_0^t dt'
\hat\psi_1^\dagger({\bf r}',t')
\hat\psi_2({\bf r}',t')
\sum_{m,n=1}^3
d_m d_n
\nonumber\\
&\times&
\left\{
\sum_{{\bf k}\lambda} 
\frac{2\pi\hbar\omega_k}{V}
e_{{\bf k}\lambda}^m
e_{{\bf k}\lambda}^n
    \exp
    \left[
         i {\bf k}
         \left(
              {\bf r} - {\bf r}'
         \right)
         - i \omega_k
         \left(
              t - t'
         \right)
    \right]
- c.c.
\right\}
\hat\psi_2({\bf r},t)
\;.
\end{eqnarray}
Introducing the commutators of the spatial components $\hat E_m({\bf r},t)$, $m=1,2,3$, of the operator of the free electric field (comp. section $C_{III}.2$ of \cite{COH89}), we may write:
\begin{equation}
\label{Xcom}
\hat X ({\bf r},t) 
=
-
\frac{i}{\hbar} 
\int d{\bf r}' \int_0^t dt'
\hat\psi_1^\dagger({\bf r}',t')
\hat\psi_2({\bf r}',t')
\sum_{m,n=1}^3
d_m d_n
\left[
     \hat{E}_m({\bf r},t), 
     \hat{E}_n({\bf r}',t')
\right]
\hat\psi_2({\bf r},t)
\;.
\end{equation}
As can be found in the section $C_{III}.3$ of \cite{COH89} the commutators may be rewritten in the following form:
\begin{eqnarray}
\label{commutator}
& &
\left[
     \hat{E}_m({\bf r},t), 
     \hat{E}_n({\bf r}',t')
\right]
\nonumber\\
&& 
=
i \hbar c
\left\{
     \left(
          \frac{3 R_m R_n}{R^2} - \delta_{mn}
     \right)
     \left[
          \frac
          {
            \delta'(R-c\tau) - \delta'(R+c\tau)
          }
          {R^2}
          -
          \frac
          {
            \delta(R-c\tau) - \delta(R+c\tau)
          }
          {R^3}
     \right]
\right.
\nonumber\\
     &&
\left.
-
     \left(
          \frac{R_m R_n}{R^2} - \delta_{mn}
     \right)
     \frac
     {
       \delta''(R-c\tau) - \delta''(R+c\tau)
     }
     {R}
\right\}
\;,
\end{eqnarray}
where ${\bf R}={\bf r}-{\bf r}'$ and $\tau=t-t'$. We introduce the polarization operator
\begin{equation}\label{def-polar}
\hat{{\bf P}}({\bf r},t) 
= 
{\bf d} \hat{\psi}_1^\dagger({\bf r},t) \hat{\psi}_2({\bf r},t) + H.c. 
= 
\hat{{\bf P}}^+({\bf r},t) + 
\hat{{\bf P}}^-({\bf r},t)
\;,
\end{equation}
and sum in (\ref{Xcom}) over $m$ and $n$ making use of (\ref{commutator}). After integration over time we get:
\begin{eqnarray}
\label{Xdf}
& &
\hat X ({\bf r},t)
=
-
{\bf d} \int d{\bf r}'
\left\{
     \frac{1}{c^2}
     \frac
     {
       \left(
           \left[
                \ddot{\hat {\bf P}}^+
           \right]
           {\bf n}
       \right)
       {\bf n}
       -
       \left[
            \ddot{\hat {\bf P}}^+
       \right]
     }
     {R}
     +
     \frac{1}{c}
     \frac
     {
       3
       \left(
           \left[
                \dot{\hat {\bf P}}^+
           \right]
           {\bf n}
       \right)
       {\bf n}
       -
       \left[
            \dot{\hat {\bf P}}^+
       \right]
     }
     {R^2}
\right.
\nonumber\\
     &&
\left.
+
     \frac
     {
       3
       \left(
           \left[
                 \hat {\bf P}^+
           \right]
           {\bf n}
       \right)
       {\bf n}
       -
       \left[
             \hat {\bf P}^+
       \right]
     }
     {R^3}
\right\}
\hat \psi_2 ({\bf r},t)
\;,
\end{eqnarray}
with ${\bf n}={\bf R}/R$ and 
$\left[\hat {\bf P}^+\right]=\hat {\bf P}^+({\bf r}',t-R/c)$. This agrees formally with an integral over the retarded fields of electric dipoles situated at $({\bf r}',t-R/c)$ (comp. section 2.2.3 of \cite{BOR68}) and may therefore be rewritten as
\begin{equation}
\label{Xrot}
\hat X ({\bf r},t) = -
{\bf d}\int d{\bf r}' \nabla \times \nabla \times 
\frac{ \left[ \hat {\bf P}^+ \right]}{R}
\hat \psi_2 ({\bf r},t)
\;,
\end{equation}
where $\nabla \times$ refers to the point ${\bf r}$.

To work out $\hat Y({\bf r},t)$ we want to introduce $\left[ {\bf P}^- \right]$ in the same way as we have introduced $\left[ {\bf P}^+ \right]$ in $\hat X({\bf r},t)$. To do so we have to change the order of the matter fields. With the help of a parameter $\varepsilon \ll t$ and with reference to the time  $\tau = t-t'$ we obtain $\hat Y({\bf r},t)$ in the form of two integrals:
\begin{eqnarray}
\label{Ylt}
\hat Y({\bf r},t) 
&=&
- i \hbar \int d{\bf r}' \int_0^\varepsilon d\tau
\sum_{{\bf k}\lambda} 
\left|
      g_{{\bf k}\lambda}
\right|^2
    \exp
    \left[
         i {\bf k}
         \left(
              {\bf r} - {\bf r}'
         \right)
         - i \omega_k \tau
    \right]
\hat\psi_2({\bf r},t)
\hat\psi_2^\dagger({\bf r}',t-\tau)
\hat\psi_1({\bf r}',t-\tau)
\nonumber\\
-
i &\hbar& \int d{\bf r}' \int_\varepsilon^t d\tau
\sum_{{\bf k}\lambda} 
\left|
      g_{{\bf k}\lambda}
\right|^2
    \exp
    \left[
         i {\bf k}
         \left(
              {\bf r} - {\bf r}'
         \right)
         - i \omega_k \tau
    \right]
\hat\psi_2^\dagger({\bf r}',t-\tau)
\hat\psi_1({\bf r}',t-\tau)
\hat\psi_2({\bf r},t)
\;.
\end{eqnarray}
Making use of the equal time commutators of the matter field $\hat \psi_2$ at $\tau \to 0$ we get
\begin{eqnarray}
\label{Ylt2}
\hat Y({\bf r},t) =
- &i& \hbar \int d{\bf r}' \int_0^t d\tau
\sum_{{\bf k}\lambda} 
\left|
      g_{{\bf k}\lambda}
\right|^2
    \exp
    \left[
         i {\bf k}
         \left(
              {\bf r} - {\bf r}'
         \right)
         - i \omega_k \tau
    \right]
\hat\psi_2^\dagger({\bf r}',t-\tau)
\hat\psi_1({\bf r}',t-\tau)
\hat\psi_2({\bf r},t)
\nonumber\\
-
&i& \hbar
\sum_{{\bf k}\lambda} 
\left|
      g_{{\bf k}\lambda}
\right|^2
\int_0^\varepsilon d\tau
    \exp
    \left(
         - i \omega_k \tau
    \right)
\hat\psi_1({\bf r},t-\tau)
\;.
\end{eqnarray}

We now turn to the sum $\hat Y({\bf r},t) + \hat Z({\bf r},t)$ and repeat the procedure we have described above when working out $\hat X({\bf r},t)$. This leads us to a term corresponding to (\ref{Xrot}) with $\left[ \hat{{\bf P}}^+ \right]$ replaced by $\left[ \hat{{\bf P}}^- \right]$ and a rest. Collecting all terms we then get
\begin{eqnarray}
\label{X+Y+Z}
&&\hat X({\bf r},t) + \hat Y({\bf r},t) + \hat Z({\bf r},t) =
- {\bf d}\int d{\bf r}' \nabla \times \nabla \times 
\frac{ \left[ \hat{{\bf P}}^+ \right] }{R}
\hat \psi_2 ({\bf r},t) 
\nonumber\\
&&
-
{\bf d}\int d{\bf r}' \nabla \times \nabla \times 
\frac{ \left[ \hat{{\bf P}}^- \right] }{R}
\hat \psi_2 ({\bf r},t)
-
i \hbar
\sum_{{\bf k}\lambda} 
\left|
      g_{{\bf k}\lambda}
\right|^2
\int_0^\varepsilon d\tau
    \exp
    \left(
         - i \omega_k \tau
    \right)
\hat\psi_1({\bf r},t-\tau)
\;.
\end{eqnarray}

We pass to the reference frame rotating with the frequency of the atomic transition $\omega_a$, so that 
$\hat \psi_2 = \hat \phi_2 \exp\left( -i\omega_a t \right)$. This shows that the term with $\left[ {\bf P}^+ \right]$ may be neglected in the rotating wave approximation. For $t\to\infty$, $\varepsilon\to\infty$ with $\varepsilon \ll t$ the last integral in (\ref{X+Y+Z}) gives
\begin{equation}
\label{gamma-delta}
- i \hbar
\sum_{{\bf k}\lambda} 
\left|
      g_{{\bf k}\lambda}
\right|^2
\left[
      \pi 
      \delta
      \left(
           \omega_k
      \right)
      - i {\cal P} 
           \left(
                \frac{1}{\omega_k}
           \right)
\right]
\hat \psi_1({\bf r},t)
\;.
\end{equation}
Calculating this term in the limit $V\to\infty$ for the quantization volume we see that it vanishes in accordance with the fact that there is no decay rate an Lamb shift for the lower atomic level. Hence (\ref{X+Y+Z}) and (\ref{heis-aa}) finally give the intended result (\ref{matter-a}) after combining the polarization term with the incident field ${\bf E}_{in}^-({\bf r},t)$ to the local field $\hat {\bf E}_{loc}^-({\bf r},t)$ according to (\ref{def-e-local}).

For the matter field $\hat \psi_2$ of the upper state we have to repeat the calculations starting with (\ref{heis-a}) and (\ref{solution-photon}) to obtain $\hat {\bf E}_{loc}^+({\bf r},t)$, correspondingly. After going to the rotating reference frame, the last integral on the r.h.s. of (\ref{X+Y+Z}) becomes in this case
\begin{eqnarray}
\label{gamma-delta2}
- i &\hbar&
\sum_{{\bf k}\lambda} 
\left|
      g_{{\bf k}\lambda}
\right|^2
\int_0^\varepsilon d\tau
    \exp
    \left(
         - i \omega_k \tau
    \right)
\hat\psi_2({\bf r},t-\tau)
\nonumber\\
=
- i &\hbar&
\sum_{{\bf k}\lambda} 
\left|
      g_{{\bf k}\lambda}
\right|^2
\int_0^\infty d\tau
    \exp
    \left[
         - i
         \left(
              \omega_k - \omega_a
         \right)
         \tau
    \right]
\hat\phi_2({\bf r},t)
\exp
\left(
     - i \omega_a t
\right)
\nonumber\\
= - i &\hbar&
\sum_{{\bf k}\lambda} 
\left|
      g_{{\bf k}\lambda}
\right|^2
\left[
      \pi 
      \delta
      \left(
           \omega_k - \omega_a
      \right)
      - i {\cal P} 
           \left(
                \frac{1}{\omega_k-\omega_a}
           \right)
\right]
\hat \psi_2({\bf r},t)
\nonumber\\
=
&\hbar& 
\left(
     \delta - i \gamma/2
\right)
\hat \psi_2 ({\bf r},t)
\;,
\end{eqnarray}
which leads to the Lamb shift $\delta$ and decay rate $\gamma$ in (\ref{matter-b}).

\section*{Appendix B. The derivation of optical Bloch equations}

In the present paper we are developing the quantum theory of light-matter interaction for second-quantized matter. It is possible to show, that the equations which are used in the usual nonlinear optics (where the matter is first-quantized) can be derived from our general equations (\ref{matter}).

Neglecting the center-of-mass motion and noise terms, we rewrite the equations (\ref{matter}) in the following manner:
\begin{mathletters}
\label{amatter}
\begin{eqnarray} 
i \hbar \frac{\partial\hat{\psi}_1({\bf r},t)}{\partial t} 
=
- 
{\bf d} 
\hat{{\bf E}}_{loc}^-({\bf r},t) \hat{\psi}_2({\bf r},t)
\;, 
\label{amatter-a}\\
i \hbar \frac{\partial\hat{\psi}_2({\bf r},t)}{\partial t} 
= 
\hbar 
\left(
     \omega_a + \delta - i \gamma/2
\right) 
\hat\psi_2({\bf r},t) -
{\bf d} \hat{\bf E}_{loc}^+({\bf r},t) 
\hat\psi_1({\bf r},t)
\;. \label{amatter-b}
\end{eqnarray}
\end{mathletters}

In the reference frame rotating with the frequency $\omega_L$ of the incident electric field the field operators have the form
\begin{equation}
\label{rrf}
\hat{\bf E}_{loc}^+
=
\hat{\bf \cal E}_{loc}^+
\exp
\left(
     - i \omega_L t
\right)
\;,
\hat\psi_2
=
\hat\phi_2
\exp
\left(
     - i \omega_L t
\right)
\;,
\end{equation}
and we obtain for (\ref{amatter})
\begin{mathletters}
\label{rrfmatter}
\begin{eqnarray} 
i \hbar \frac{\partial\hat{\psi}_1({\bf r},t)}{\partial t} 
=
- {\bf d} 
\hat{\bf \cal E}_{loc}^-({\bf r},t) \hat{\phi}_2({\bf r},t)
\;, 
\label{rrfmatter-a}\\
i \hbar \frac{\partial\hat{\phi}_2({\bf r},t)}{\partial t} 
= 
- \hbar 
\left(
     \Delta + i \gamma/2
\right) 
\hat\psi_2({\bf r},t) -
{\bf d} \hat{\bf \cal E}_{loc}^+({\bf r},t) 
\hat\psi_1({\bf r},t)
\;, 
\label{rrfmatter-b}
\end{eqnarray}
\end{mathletters}
with the detuning $\Delta=\omega_L-\omega_a+\delta$.

In the rotating reference frame the macroscopic polarization operator has the form
\begin{equation}
\hat {\bf P}^+ 
=
{\bf d} \hat \rho \hat R^+
\exp
\left(
     - i \omega_L t
\right)
\;,
\end{equation}
where $\hat \rho \hat R^+ = \hat \psi_1^\dagger \hat \phi_2$ and 
$\hat \rho = \hat \psi_1^\dagger \hat \psi_1 + \hat \phi_2^\dagger \hat \phi_2$ is the total density of the atoms which is a sum of the densities of the atoms in the ground and the excited state.

Making use of the eqs. (\ref{rrfmatter}) one can show that the time evolution of the operator $\hat R^+$ is described by the following equation
\begin{equation}
\label{Reqn}
\frac{\partial \hat R^+}{\partial t}
=
\left(
     i \Delta - \gamma/2
\right)
\hat R^+
-
\frac{i}{\hbar}
{\bf d} \hat {\bf \cal E}_{loc}^+ \hat W
\;,
\end{equation}
where
\begin{equation}
\label{inversion}
\hat W = 
\left(
     \hat \phi_2^\dagger \hat \phi_2 - \hat \psi_1^\dagger \hat \psi_1
\right)
/\hat \rho
\end{equation}
is the inversion operator. Its time evolution is governed by the Heisenberg equation:
\begin{equation}
\label{Weqn}
\frac{\partial \hat W}{\partial t}
=
-\frac{\gamma}{2}
(1+\hat W)
+
\frac{2i}{\hbar}
\left[
    {\bf d} \hat {\bf \cal E}_{loc}^+ \hat R^-
    -
    {\bf d} \hat R^+ \hat {\bf \cal E}_{loc}^-
\right]
\;.
\end{equation}

In the derivation of the equations (\ref{Reqn}), (\ref{Weqn}) we have taken into account that the total density of the atoms $\rho$ is an integral of motion. It is possible to show this starting with the second-quantized Hamiltonian (\ref{ham-second-quant}) where the first term, which describes the center-of-mass motion, is neglected. However, it is not possible to show this when starting with the equations (\ref{amatter}) because in these equations the vacuum fluctuations of the photon field are neglected. The neglect of the vacuum fluctuations in the equations (\ref{amatter}) corresponds to the neglect of some terms which are not presented explicitly in the Hamiltonian (\ref{ham-second-quant}).

Replacing the spontaneous emission rate $\gamma/2$ in the equations (\ref{Reqn}), (\ref{Weqn}) by the phenomenological damping and relaxation terms, $\gamma_L$ and $\gamma_T$, respectively, we get the optical Bloch equations\cite{ALL78,BOW93}. 


\end{document}